%% file: DyonsOnSasakianMfs.tex
\documentclass[12pt]{article}	

\usepackage[latin1]{inputenc}	
\usepackage{graphicx}
\usepackage{color}
\usepackage[sort&compress,square,comma,numbers]{natbib}
\usepackage[usenames,dvipsnames,svgnames,table]{xcolor}
\usepackage{amsmath}
\usepackage{amssymb}
\usepackage{array}
\usepackage{multirow}
\usepackage{amsthm}
\usepackage{bbm}
\usepackage{verbatim}
\usepackage{cancel}
\usepackage{tabularx}
\usepackage[all]{xy}
\usepackage[colorlinks, citecolor=blue, linkcolor=blue, urlcolor=Maroon, filecolor=Maroon]{hyperref}

\numberwithin{equation}{section} 
\def\fnote#1#2{\begingroup\def\thefootnote{#1}\footnote{#2}
     \addtocounter{footnote}{-1}\endgroup}

\include{Pre}

\begin{document}

\begin{titlepage}

\vfill
\begin{flushright}
\normalsize{ITP--UH--25/14}
\end{flushright}

\vfill

\begin{center}
   \baselineskip=16pt
   {\Large \bf Yang-Mills Solutions and Dyons on Cylinders \\ \vspace{10pt} over Coset Spaces with Sasakian Structure}
   \vskip 2cm
   Maike Torm\"ahlen
   \vskip .6cm
   \begin{small}
    {\it Institut f\"ur Theoretische Physik,\\
         Leibniz Universit\"at Hannover, Appelstra\ss e 2, 30167 Hannover, Germany}
   \end{small}
\end{center}

\vfill 
\begin{abstract}
\noindent
We present solutions of the Yang-Mills equation on cylinders $\bR\times G/H$ over coset spaces of odd dimension $2m+1$ with Sasakian structure. The gauge potential is assumed to be $SU(m)$-equivariant, parametrized by two real, scalar-valued functions. Yang-Mills theory with torsion in this setup reduces to the Newtonian mechanics of a point particle moving in $\bR^2$ under the influence of an inverted potential. We analyze the critical points of this potential and present an analytic as well as several numerical finite-action solutions. Apart from the Yang-Mills solutions that constitute $SU(m)$-equivariant instanton configurations, we construct periodic sphaleron solutions on $S^1\times G/H$ and dyon solutions on $i\bR\times G/H$.
\end{abstract}

\vfill
\setcounter{footnote}{0}
\setcounter{page}{0}

\fnote{}{maike.tormaehlen@itp.uni-hannover.de}

\end{titlepage}

\tableofcontents

\bigskip
\section{Introduction}
Higher-dimensional Super-Yang-Mills theory appears in the context of string theory for example as the low-energy limit of the heterotic superstring. In this limit, heterotic string theory yields ten-dimensional supergravity coupled to $\cn=1$ supersymmetric Yang-Mills theory \cite{Green1987_1, Blumenhagen2007}. In four dimensions, the full Yang-Mills equation is implied by the instanton equation, a first-order anti-self-duality equation. This fact generalizes to dimensions greater than four. The higher-dimensional instanton equation is particularly interesting for string compactifications on manifolds of the form $M_{10-d}\times X^d$ with compact part $X^d$ and maximally symmetric flat space $M_{10-d}$. Requiring the gauge field on the compact manifold to satisfy the instanton equation is closely related to the preservation of $\cn=1$ supersymmetry on the non-compact spacetime part. Instantons in higher dimensions were first studied in \cite{Corrigan1983}, and solutions to the generalized anti-self-duality equation have been constructed, for example, in \cite{Fairlie1984a, Fubini1985, Ivanova2008, Ivanova2009, Harland2010, Haupt2011, Gemmer2011}.
\bigskip

\newpage
The requirement of supersymmetry preservation in string compactifications of the above type translates to a condition on the geometry of the compact internal manifold: imposing the instanton equation on the gauge field on $X^d$ is equivalent to requiring reduced holonomy on the compact space. In heterotic compactifications, Calabi-Yau 3-folds have therefore been the preferred choice for compactification spaces, leading to phenomenologically interesting mo\-dels with $\cn=1$ supersymmetry. Furthermore, $G_2$-holonomy 7-manifolds as well as 8-manifolds with $Spin(7)$-holonomy have been of interest in more general models. A problem of heterotic Calabi-Yau compactifications is that they come with a number of scalar fields with undetermined vacuum expectation value. Some of these moduli can be fixed by allowing for nonvanishing $p$-forms, so-called fluxes, to exist on the internal compact manifold. Flux compactifications do address the moduli problem but enlarge the number of possible string backgrounds significantly, leading to the string landscape problem. For a review of flux compactifications, see for example \cite{Grana2006, Blumenhagen2006, Douglas2006}.
\bigskip

Nontrivial background fluxes on the internal compact manifold imply a backreaction on the geometry, relaxing the condition on the holonomy of the manifold. $X^d$ is no longer required to have reduced holonomy but to admit a $G$-structure, i.e. a reduction of the tangent bundle structure group from $GL(d)$ to some subgroup $G\subset GL(d)$. If the manifolds admit a real Killing spinor \cite{Baer1993}, they are equipped with a connection with totally antisymmetric torsion. In the following, we will be interested in a connection whose torsion is determined up to a real scaling parameter $\k$. We will consider Sasakian manifolds as a special type of Killing spinor manifolds of dimension $2m+1$ with structure group $G=SU(m)$. For a particular choice of metric, these manifolds are in addition Einstein. Sasaki-Einstein manifolds have been studied in the context of non-compact flux backgrounds as AdS/CFT duals of confining gauge theories or, more precisely, as type IIB AdS vacua that lead to dual $\cn=1$ Super Yang-Mills theories coupled to matter \cite{Grana2006, Cassani2010, Gauntlett2010, Gauntlett2006}. 
\bigskip

In this paper, we concentrate on cylinders $\bR\times G/H$ over coset spaces with Sasakian structure. We start by repeating the basics of Yang-Mills equations, $G$-structure and in particular Sasakian manifolds in Chapter \ref{sec:SasakiMfs}. We use an $SU(m)$-equivariant ansatz for the gauge connection, parametrized by two real scalar functions, to write out the Yang-Mills equation in components in Chapter \ref{ch:YMonSE}. This leads to a system of two coupled second-order ordinary differential equations, reducing Yang-Mills theory with torsion to the Newtonian mechanics of a point particle moving in $\bR^2$ under the influence of a potential. The shape of this potential depends on the torsion parameter $\k$. The instanton case is recovered for $\k=1$. This case has been first studied in \cite{Correia2009, Correia2010} and can also be found in \cite{Harland2011}. We derive the corresponding particle action in Chapter \ref{ch:Action} and discuss the critical points of zero energy. For a special value of $\k$, the second-order equations can be solved analytically and yield a $tanh$-kink-type solution, similar to solutions discussed in earlier works \cite{Rahn2009, Bauer2010}. We construct further finite-action solutions numerically. Considering $S^1\times G/H$ instead of $\bR\times G/H$, we obtain periodic solutions, so-called sphalerons, which are discussed in section \ref{ch:PeriodicSolutions}. Taking the product space $i\bR\times G/H$ instead of $\bR\times G/H$ leads to a sign flip in the potential. Solutions to this case are known as dyons and can be constructed numerically. We present some of them in section \ref{ch:DyonSolutions}.

\section{Sasakian manifolds}
\label{sec:SasakiMfs}
As described in the introduction, compact $G$-structure manifolds play a key role in string compactifications. Particularly interesting in this context are real Killing spinor manifolds (see also \cite{Ivanova2012} for details). Besides the round spheres, these are 
\begin{itemize}
	\item 6-dimensional $SU(3)$-structure manifolds,
	\item 7-dimensional $G_2$-structure manifolds,
	\item $(2m{+}1)$-dimensional Sasakian manifolds with structure group $SU(m)$,	
	\item $(4m{+}3)$-dimensional 3-Sasakian manifolds with structure group $Sp(m)$.
\end{itemize}
In this note, we consider Sasakian manifolds of dimension $2m+1$, where $1\leq m\in \bN$. A detailed introduction to Sasakian geometry can be found in \cite{Sparks2010} and \cite{Boyer2008}. Let us review the most important facts here. 
\bigskip

Sasakian manifolds are special types of contact manifolds. According to \cite{Andrada2008, Conti2007}, an \textit{almost contact structure} $(\Phi,\eta,\xi)$ on an odd-dimensional Riemannian manifold $(M,g_M)$ is characterized by a nowhere vanishing vector field $\xi$ and a one-form $\eta$, satisfying $\eta(\xi)=1$, plus a $(1,1)$-tensor $\Phi$ such that $\Phi^2=-\id + \xi\otimes \eta$. An almost contact structure is called \textit{contact} if in addition the one-form satisfies
\begin{align}
	\eta\wedge (d\eta)^m \neq 0.
\end{align}
In this case, $\eta$ is called \textit{contact form}, and $\xi$ is referred to as \textit{Reeb vector field}. Contact structures are \textit{normal} if for their Nijenhuis tensor $N$ associated to $\Phi$,
\begin{align}
	N_\Phi(X,Y) = \Phi^2[X,Y] + [\Phi X,\Phi Y] - \Phi[\Phi X,Y] - \Phi &[X,\Phi Y] 
	\ \forall\ X,Y\in \G(TM),
\end{align}
the relation $N=-d\eta\otimes \xi$ holds.%
\footnote{This is equivalent to the complex structure $J$ induced on the product manifold $\bR\times M$ being integrable.}
When the Riemannian metric $g$ on an almost contact manifold $(M,g)$ satisfies 
\begin{align}
	g_M(\Phi X,\Phi Y) = g_M(X,Y)-\eta(X) \eta(Y)
\end{align}
for any two vector fields $X,Y\in TM$, the structure is referred to as \textit{almost contact metric}. It is called \textit{contact metric} if in addition 
\begin{align}
	d\eta=2\o, \label{eq:Sasaki}
\end{align}
with a two-form $\o(X,Y):=g(X,\Phi Y)$, is satisfied.
\bigskip

A \textbf{Sasakian manifold} is defined to be a manifold with \textbf{normal contact metric structure}. Such manifolds admit a reduction of the tangent bundle structure group from $SO(2m+1)$ to $U(m)$, which allows (apart from the existence of the one-form $\eta\in\O^1(M)$ and two-form $\o\in\O^2(M)$) for the introduction of invariant forms $P_M\in\O^3(M)$ and $Q_M\in\O^4(M)$ that satisfy the following relations:
\begin{align}
	P_M &= \h\La \o, && Q_M = \frac{1}{2} \o\La \o, && \h \lrcorner \o = 0. \label{eq:PQ}
\end{align}
The contraction is defined as $\h\lrcorner \o=*_M(\h\La*_M\o)$ by use of the Hodge star operator on $(M,g_M)$ (see for example \cite{Harland2010}). All these forms are parallel with respect to the canonical connection introduced below. In addition to equation (\ref{eq:Sasaki}), the forms satisfy the relations
\begin{align}
	d*_M\o &= 2m*_M\h,\\
	dP_M	&= 4Q_M, \\ 
	d*_M Q_M &= (2m-2) *_M P_M.
\end{align}
Condition (\ref{eq:Sasaki}) can be generalized. If the structure satisfies $d\h=\a\o$ for some real $\a$, it is referred to as $\a$-Sasakian. We will see below that in our case the $\a$-Sasakian structure can be transformed into a Sasakian structure by rescaling of basis elements.
\bigskip

If the metric on a Sasakian manifold is proportional to the Ricci tensor, $Ric\propto g$, we have a \textbf{Sasaki-Einstein manifold}. Note that a Sasakian structure need not necessarily be Einstein. Sasaki-Einstein manifolds admit a reduction of the tangent bundle structure group to $SU(m)$. A comment on the relation of the structure groups of Sasakian and Sasaki-Einstein manifolds will be given in Chapter \ref{ch:LieAlgStr}.

\vspace{1cm}
\subsection{Canonical connection}
Connections on the tangent bundle $TM$ over a manifold $M$ are locally determined by matrix-valued one-forms $\G_\mu^\nu=\G_{\s\mu}^\nu e^\s$, using a basis $\{e^\s\}$ of non-holonomic one-forms and Greek indices $\mu = \{1,2,...,2m+1\}$ to label directions on $M$. We introduce the so-called \textit{canonical} connection on the tangent bundle of a $G$-structure manifold $M$ as a connection with holonomy group $G$ and totally antisymmetric torsion with respect to some $G$-compatible metric \cite{Harland2011}. We denote this connection by $\n^P$, or, locally, by $^P\G_\mu^\nu$. All of the above listed real Killing spinor manifolds come equipped with canonical 3- and 4-forms $P_M$ and $Q_M$, and in all these cases the torsion of the canonical connection is proportional to the 3-form $P_M$ \cite{Harland2011}. The canonical connection can therefore be constructed as a sum of the (torsion-free) Levi-Civita connection $\n^{LC}$ and the 3-form $P_M$, and is in the Sasakian case given by 
\begin{align}
	^P\G_{\mu a}^b &= ^{LC}\G_{\mu a}^b + \frac{1}{m} P_{\mu ab}, \label{eq:Nabla_P_Sasaki_1} \\
	^P\G_{\mu 1}^a &= -^P\G_{\mu a}^1 = ^{LC}\G_{\mu 1}^a + P_{\mu 1a}. \label{eq:Nabla_P_Sasaki_2}
\end{align}
Another connection in the tangent bundle over $M$ will play a special role in the following. The coefficients of any metric-compatible torsionful connection $\G$ on $TM$ are uniquely determined by the conditions 
\begin{align}
	dg_{\mu \nu}-g_{\mu \r}{\G^\r}_\nu-g_{\nu\r}{\G^\r}_\mu&=0, \label{metric} \\
	d e^\mu+\G_{\nu\r}^\mu e^\nu\wedge e^\r &=T^\mu, \label{torsion}
\end{align}
where $T^\mu=\frac{1}{2}T_{\nu\r}^\mu e^\nu\wedge e^\r$ denotes the torsion two-form. Motivated by its appearance in heterotic supergravity, we introduce the \textbf{torsionful spin connection} $\G$ with components 
\begin{align}
	\G_{\mu\nu}^\r = \ ^{LC}\G_{\mu\nu}^\r + T_{\mu\nu}^\r
\end{align}
as a metric-compatible connection with totally antisymmetric torsion. This connection will be used later on for the construction of Yang-Mills solutions, and we will explicitly compute its coefficients. The torsion is chosen to be proportional to the structure constants, $T_{\mu\nu\r}\propto f_{\mu\nu\r}$. 
We use small indices $a = \{2,\ldots,2m+1\}$ for directions on $M$ excluding the contact direction 1. It is useful to choose a local orthonormal basis $\{e^1,e^a\}$ of $T^\ast M$ such that the parallel forms become 
\begin{align}
	\h &= e^1, && \o = e^{23} + e^{45} + \cdots + e^{2m, 2m{+}1}. \label{eq:StandardForm}
\end{align}
Here and in the following, we use the shorthand notation $e^a\La e^b=e^{ab}$. The canonical connection $\n^P$ is compatible with the following family of metrics, all of which are Sasakian up to homothety:
\begin{align}
	g_h &= e^1 e^1 + e^{2h} \d_{ab} e^a e^b. \label{eq:SEMetric}
\end{align}
This can be seen by rescaling the metric with a real parameter $\g$,
\begin{align}
	g_{h,\g} &= \g^2(e^1 e^1 + e^{2h} \d_{ab} e^a e^b),
\end{align}
and introducing new basis forms $\wt e^1 = \g e^1, \wt e^a = \g e^h e^a$, such that the metric takes the form
\begin{align}
	g_{h,\g} &= \wt e^1 \wt e^1 + \d_{ab} \wt e^a \wt e^b.
\end{align}
Recall that the original basis one-forms satisfy the contact condition $d e^1=2\o$. For still being Sasakian after rescaling, the new structure constants have to satisfy an analogous condition. We find
\begin{align}
 d \wt e^1 = \frac{2}{\g e^{2h}}\wt \o := \a(h,\g)\wt \o.
\end{align}
The structure is therefore $\a$-Sasakian for all $\a(h,\g)$ (hence also for all scaling factors $\g$) and Sasakian (i.\ e.\ $d\wt e^1=2\wt \o$) for the special value $\a=2$, or, equivalently, $\g=e^{-2h}$.

\bigskip
For $h=0$, the structure becomes Einstein. The value $e^{2h}=\frac{2m}{m+1}$ is special, as it makes the torsion of the canonical connection totally antisymmetric. Connections with totally antisymmetric torsion are particularly interesting for string compactifications. We will therefore restrict our consideration to the latter case from now on and not study the Einstein case in detail.

\subsection{Lie algebra structure and connection on Sasakian manifolds}
\label{ch:LieAlgStr}
As mentioned in Chapter \ref{sec:SasakiMfs}, Sasakian manifolds admit a reduction of the tangent bundle structure group to $U(m)$, while the structure group on Sasaki-Einstein manifolds reduces further to $SU(m)$ (see also \cite{Boyer2008}). We consider a Sasakian manifold $M$ with metric (\ref{eq:SEMetric}), which is not Einstein for arbitrary values of the parameter $h\in\bR$ but allows for an $SU(m)$ structure. To understand this construction, first note that there is no one-to-one correspondence of the existence of an $SU(m)$ structure on $M$ and the Einstein property of the metric. If a manifold is Sasaki-Einstein, it must have $SU(m)$ structure group, but the converse is not necessarily true.

\bigskip
The structure of our manifold arises as follows. Starting with a Sasaki-Einstein manifold that has structure group $SU(m)$ and admits two Killing spinors $\e$, $\tilde\e$, one may construct the canonical connection $\n^P$ by the requirement $\n^P\e=\n^P\tilde \e=0$. $\n^P$ has holonomy $SU(m)$ and components (\ref{eq:Nabla_P_Sasaki_1}), (\ref{eq:Nabla_P_Sasaki_2}). As $\n^P$ is compatible with the whole family of metrics (\ref{eq:SEMetric})%
\footnote{The compatibility can be verified by explicitly computing $\n^P g_h$, using $g_h$ to raise and lower indices, or by rewriting $g_h$ in terms of $e^1$ and $g_{h=1}$ and employing $\n^P e^1=\n^P g_{h=1}=0$. We thank Derek Harland for this comment.},
deformation of the metric does not affect the spinor identities $\n^P\e = \n^P\tilde \e = 0$. The existence of two Killing spinors, on the other hand, is in one-to-one correspondence with the existence of an $SU(m)$-structure. Hence, the family $g_h$ of metrics preserves the $SU(m)$ structure although the Einstein property is lost after deformation of the metric. We therefore have a Sasakian manifold with $SU(m)$-structure that is explicitly not Einstein.

\bigskip
We will consider Yang-Mills theory on the cylinder $\cz(M)=\bR\times M$ over a Sasakian coset space $M=G/H$ with structure group $SU(m)$%
\footnote{A discussion about the Yang-Mills equation on Sasakian manifolds that do not necessarily have coset structure can be found in \cite{Tormaehlen2015}.}. 
The metric on $\cz(M)$ is given by
\begin{align}
	g &= e^0 e^0 + e^1 e^1 + \frac{2m}{m+1} \d_{ab} e^a e^b, \label{eq:SEMetric2a}
\end{align}
where $e^0:=d\t$ denotes the coordinate in $\bR$ direction. The canonical connection on $M$ lifts to a connection on the tangent bundle over $\cz(M)$ with holonomy group $SU(m)$. We will introduce a more general connection on the tangent bundle $T\cz(M)$, namely a perturbation of $\n^P$ by parallel sections (cf. \cite{Harland2011} for details). The holonomy group of this perturbed connection is $SU(m+1)$. This corresponds to having an $SU(m+1)$-structure on $\cz(M)$ and hence to a principal bundle $P(\cz(M),SU(m+1))$. The existence of an $SU(m)$-holonomy connection in $T\cz(M)$ furthermore implies the existence of a principal subbundle of $P$ with structure group $SU(m)$. As $SU(m)$ is a subgroup of $SU(m+1)$, the corresponding Lie algebras split according to $\mathfrak{su}(m+1)=\mathfrak{su}(m)\oplus\gm$, where $SU(m)$ acts irreducibly on $\mathfrak{su}(m)$ and $\gm$ denotes the $(2m+1)$-dimensional orthogonal complement to $\mathfrak{su}(m)$. Representation theoretic arguments and the requirement of $SU(m)$-equivariance (see for example \cite{Bauer2010, Tormaehlen2015} for details) then allow for a connection of the following form, where $\chi(\t)$ and $\psi(\t)$ are real functions of the variable $\t$ parametrizing the cylinder direction:
\begin{align}
	\ca = \n^P + \chi(\t)e^1 I_1 + \frac{1}{\sqrt{2m}} \psi(\t) e^a I_a. \label{eq:A}
\end{align}
We denote the generators of $\mathfrak{su}(m)$ as $\{I_i\}$, the dual one-forms by $\{e^i\}$ and the generators of $\gm$ as $\{I_1,I_a\}$. They are chosen to be dual to the above introduced one-forms $\{e^1,e^a\}$. The frame $\{e^i\}$ on $\gs\gu(m)$ can be expressed as a linear combination of the one-forms $\{e^\mu\}$ as $e^i=e^i_\mu e^\mu$, where the $e^i_\mu$ are real functions. Written as matrices, the generators have the following nonvanishing entries:
\begin{align}
	I_{ia}^b &= f_{ia}^b, \nn \\
	I_{1a}^b &= -\frac{1}{m} \o_{ab}, && -I_{11}^0 = I_{10}^1 = 1, \label{eq:Generators} \\
	-I_{ab}^0 &= I_{a0}^b = \d_a^b, && -I_{ab}^1 = I_{a1}^b = \o_{ab}. \nn
\end{align}
In this basis, the structure constants satisfy
\begin{align}
	f_{ab}^1 &= 2 P_{ab1}, && f_{1a}^b = \frac{m+1}{m} P_{1ab}. \label{eq:StrConst}
\end{align}
Note that these identities differ from the corresponding equations in \cite{Harland2011} by a sign. The sign here is in agreement with equations (\ref{eq:Generators}).

\vspace{1cm}
\section{Yang-Mills equation on Sasakian manifolds}
\label{ch:YMonSE}
Before specializing to Yang-Mills theory on Sasakian coset spaces, let us review some general facts about the Yang-Mills equation. Consider a Riemannian manifold $(N,g_N)$ of dimension $d>4$, and let $E$ be a complex vector bundle over $N$, endowed with a connection $\ca$. We denote by $\cf=d\ca+\ca\La \ca$ the curvature of this connection and by $*$ the Hodge star operator with respect to the metric $g_N$. The generalized anti-self-duality equation is well-defined on any manifold equipped with an invariant 4-form $Q\in \O^4(N)$, in particular on any $G$-structure manifold, and reads 
\begin{align}
	*\cf = -*Q\La \cf.
\end{align}
Applying the gauge covariant derivative $\cd=d+[\ca,\cdot]$ leads to the torsionful Yang-Mills equation
\begin{align}
	d*\cf + [\ca,*\cf] + *\ch\La \cf = 0 \label{eq:YMTorsion1},
\end{align}
where we have defined a 3-form $\ch$ via $*\ch:=d*Q$. If $d*Q\La \cf=0$, this equation reduces to the standard Yang-Mills equation without torsion. The torsionful Yang-Mills equation is the equation of motion for the following action, which includes the standard Yang-Mills plus a Chern-Simons term:
\begin{align}
	S = \int_M \tr\left(\cf\La *\cf + (-1)^{d-3} *Q\La\cf\La\cf\right). \label{eq:S_YM_CS}
\end{align}
The torsion term in the Yang-Mills equation is generated by variation of the Chern-Simons term, while the other summands arise from variation of the Yang-Mills term. 

\bigskip
For a motivation of the following discussion, recall that a connection with totally antisymmetric torsion naturally appears in the conditions for supersymmetry preservation in heterotic supergravity \cite{Becker2007}. On suitably chosen string backgrounds, one can introduce geometric three-form fluxes that are identified with the torsion of this spin connection. The Yang-Mills equation follows from covariant differentiation of the higher-dimensional instanton equation if the three-form is related to the $G$-structure four-form as $*\ch:=d*Q$. In this case, the Yang-Mills equation is the equation of motion of the action (\ref{eq:S_YM_CS}). 

\bigskip
Non-BPS Yang-Mills solutions can be constructed when the Yang-Mills equation is not required to follow from a first-order equation. In accordance with earlier work \cite{Ivanova2009, Harland2010, Bauer2010}, we choose to identify the three-form $\ch$ with the torsion of the spin connection, $\ch_{ABC}\propto T_{ABC}$, and the torsion components%
\footnote{It has been argued in \cite{Harland2010} that for such a choice of $\ch_{ABC}$ and $T_{ABC}$, the Yang-Mills equation on the cylinder over a nearly-Kähler coset space follows from an action similar to (\ref{eq:S_YM_CS}). This does not have to hold for other choices of $\ch$.}
to be proportional to the structure constants on $G/H$:
\begin{align}
	T_{abc} = \k f_{abc}, && \k \in \bR. \label{eq:TorsionKappa}
\end{align}
In explicit examples, the relation of $T$ and $\ch$ will be chosen such that $*\ch=d*Q$ is satisfied for $\k=1$ and the Yang-Mills equation follows from the instanton equation for this value of $\k$. Other choices are possible and correspond to a rescaling of the parameter $\k$. Solutions of the torsionful Yang-Mills equation can be lifted to solutions of heterotic supergravity if they follow from a first-order BPS equation. More general non-BPS Yang-Mills solutions for arbitrary values of $\k$ can potentially serve as building blocks for non-supersymmetric string solutions.

\bigskip
Written out in components, the torsionful Yang-Mills equation on the product space $\bR\times G/H$ turns into the following set of equations, where the metric $g$ is assumed to be of diagonal form with coordinate-dependent components:
\begin{align}
	\frac{g_{BB}}{\sqrt{|g|}} \p_C& \left( \sqrt{|g|} \cf^{CB} \right) \nn \\
		&- \cf^{CD} \left(\frac{1}{2}{T_{CDB}} - {\G_{CDB}}\right) + {\cf^C}_B\left(\frac{1}{2}{T_{CD}}^D-{\G_{CD}}^D \right) \nn \\
		&- {\cf^C}_B\left(\frac{1}{2}{T_{DC}}^D-{\G_{DC}}^D\right) + [\ca^A, F_{AB}] - \frac{1}{2} {\ch_{CDB}} \cf^{CD} = 0. \label{eq:YMComp}
\end{align}
We study the Yang-Mills equation on the cylinder $\cz(M) = \bR\times M$ with metric (\ref{eq:SEMetric2a}), where $M=G/H$ is a coset space with Sasakian structure of dimension $2m+1$. In this setup, $G$ is a semisimple Lie group and $H$ a closed Lie subgroup. Capital indices $A = \{0,1,2,...,2m+1\}$ are used to label all $2m+2$ directions of the product manifold. The upper index will always be pulled down \textit{behind} the lower two ones. This convention is important, as not all quantities do a priori have totally antisymmetric indices. 

\bigskip
As the free index $B$ runs from 0 to $\dim G/H$, these are $2m+2$ equations. The coefficients $\ch_{ABC}$ (with all indices lowered) are the components of the 3-form $\ch$, and $\ ^-\G_{AB}^C$ are the coefficients of the torsionful spin connection with torsion $T_{AB}^C$. Equation (\ref{eq:YMComp})%
\footnote{Note that this equation is not identical to the corresponding equations (2.19) and (2.20) in \cite{Bauer2010} due to differently normalized torsion. The equations presented in the reference follow from our equation (\ref{eq:YMComp}) with cylinder metric $g_\cz = d\t^2 + \d_{ab}e^a e^b$ in the special case of $H_{ABC}=-T_{ABC}$.
} 
has been discussed in detail on the cylinder over an arbitrary coset space $G/H$ with gauge connection $\ca=e^i I_i + \phi e^a I_a$ in \cite{Bauer2010, Ivanova2009, Rahn2009, Harland2010}, leading to explicit kink-type solutions.

\bigskip
For further specification, we have to compute the components of the 3-form $\ch$. According to \cite{Harland2011}, an invariant 4-form on the cylinder over a Sasakian manifold can be constructed as 
\begin{align}
	Q = \frac{2m}{m+1} d\t\La P_M + \left(\frac{2m}{m+1}\right)^2 Q_M.
\end{align}
A direct computation yields
\begin{align}
	*\ d*Q &= -\frac{5}{2\cdot 3!} {f^{KL}}_{[R}Q_{KLPQ]} e^{PQR}.
\end{align}
Using the definition $*\ch:=d*Q$ and decomposition rules for antisymmetric tensor indices, we find
\begin{align}
	\ch_{PQR} &= -\frac{5}{2} {f^{KL}}_{[R}Q_{KLPQ]}
		= -\frac{15}{2} Q_{KL[PQ}{f^{KL}}_{R]}.
\end{align}
At this point, we have to distinguish between indices in cylinder direction $(0)$, contact direction $(1)$ and all other directions and find that the following components vanish for all $m$:
\begin{align}
	\ch_{01r} &= \ch_{0qr} = \ch_{pqr} = 0.
\end{align}
The remaining components depend on the value of $m$. We demonstrate this by writing out $\ch_{231}$ explicitly, using equations (\ref{eq:PQ}), (\ref{eq:StandardForm}) and (\ref{eq:StrConst}). All other nonvanishing components of $\ch$ behave in a similar way:
\begin{align}
\begin{array}{l c l}
	\ch_{231} = -2 P_{231} Q_{2323} = 0 && \text{ for } m=1, \vspace{10pt} \\
	\ch_{231} = -\dfrac{9}{16} P_{mn1} Q_{mn23} = -2 P_{451} = -f_{231} && \text{ for } m=2, \vspace{10pt} \\
	\ch_{231} = -\dfrac{4}{9} P_{mn1} Q_{mn23} = -2 (P_{451}+P_{671}) = -2 f_{231} && \text{ for } m=3, \vspace{10pt} \\
	\ch_{231} = (1-m) f_{231} && \text{ for arbitrary } m. \vspace{10pt} \\
\end{array} \label{eq:HT}
\end{align}
Note that the case $m=1$ of lowest dimension with $\ch=0$ is special. We will not further discuss it here. In order to recover the instanton case for $\k=1$, we choose 
\begin{align}
	H_{\mu\nu\r} = (1-m) T_{\mu\nu\r} = (1-m) \k f_{\mu\nu\r}.
\end{align}
With this choice and the cylinder metric (\ref{eq:SEMetric2a}), equation (\ref{eq:YMComp}) turns into
\begin{align}
	\p_A \cf^{AB} &- \cf^{CD}\left(\frac{1}{2}(2-m)T_{CD}^{B}-\G_{CD}^{B}\right) \nn \\
		&+ \cf^{CB}\left(\frac{1}{2}T_{CD}^D-\G_{CD}^{D}\right) - \cf^{CB}\left(\frac{1}{2}T_{DC}^D-\G_{DC}^{D}\right) + [\ca_A, \cf^{AB}] = 0, \label{eq:YMTorsion}
\end{align}
where $\G$ denotes the torsionful spin connection. The $B=0$ equation is identically satisfied. Let us take a look at the cases with  $B>0$. The summand $\cf^{C\mu}(\frac{1}{2}T_{CD}^D-\G_{CD}^{D})$ vanishes identically. From the summand $\cf^{C\mu}(\frac{1}{2}T_{DC}^D-\G_{DC}^{D})$, as well as from $[\ca_A, \cf^{AB}]$, we obtain terms proportional to the functions $e_\mu^i$. These terms add up to zero by use of the Jacobi identity and $SU(m)$-equivariance of the connection and will therefore be omitted in the following computation. We evaluate the remaining terms explicitly. The connection coefficients are derived from the Maurer-Cartan structure equation
\begin{align}
	T^A = d e^A + \G_{BC}^A e^{BC}.
\end{align}
With
\begin{align}
	de^A &= -\frac{1}{2}f_{BC}^A e^{BC} && \text{and} & T_{BC}^A = \k f_{BC}^A, \label{eq:Torsion}
\end{align}
where $\k\in \bR$ is a real parameter, they take the form
\begin{align}
	\G_{bc}^1 &= \frac{1}{2} (\k+1) f_{bc}^1, \\
	\G_{1b}^a &= \frac{1}{2} (\k+1) f_{1b}^a + f_{ib}^a e^i_c, \\
	\G_{bc}^a &= f_{ic}^a e^i_b.
\end{align} 
By construction, the value $\k=1$ describes the instanton case discussed in \cite{Correia2009, Correia2010, Harland2011}. For the derivation of explicit second-order equations, we will use 
\begin{align}
	T^1 &= P_{1ab} e^{1ab}, && T^a = \frac{m+1}{2m} P_{a\mu \nu} e^{a \mu \nu},
\end{align}
along with equations (\ref{eq:StrConst}). Using equation (\ref{eq:A}) and omitting the $\t$-dependence of the functions $\chi$ and $\psi$, we obtain the following curvature:
\begin{align}
	\cf = &- \frac{1}{2} \left(1-\frac{1}{2m}\psi^2\right) f_{ab}^i e^{ab} I_i 
				 + \dot \chi e^{01} I_1 + \frac{1}{\sqrt{2m}} \dot \psi e^{0a} I_a \nn \\
				&+ \left(\chi-\frac{1}{2m}\psi^2\right) P_{ab1} e^{ab} I_1 + \frac{m+1}{m\sqrt{2m}} \psi(1-\chi) P_{1ba} I_a e^{1b}. \label{eq:FSasaki}
\end{align}
Inserting $\cf$, $T^A$, $\o_{AB}^C$ as above and using 
\begin{align}
	f_{ac}^i f_{ib}^c &= \frac{2 (m^2-1)}{m} \d_{ab}, \label{eq:beta}
\end{align}
equation (\ref{eq:YMTorsion}) turns into
\begin{subequations}
\begin{align}
	\ddot \chi &= \frac{(m+1)^2}{m} \left(\left((m-1)\k+1\right)\chi - \left((m-1)\k+3\right) \frac{1}{2m}\psi^2 + \frac{1}{m} \chi\psi^2 \right), \\
	\ddot \psi &= \left(\frac{m+1}{m}\right)^2 \psi \left((m-1)\k+2-m - \left((m-1)\k+3\right) \chi + \chi^2 + \frac{1}{2}\psi^2 \right).
\end{align} \label{eq:YMSE1}
\end{subequations}
The derivation of the identity (\ref{eq:beta}) can be found in Appendix \ref{sec:beta}.

\section{Action functional and potential}
\label{ch:Action}
The second-order equations (\ref{eq:YMSE1}) are equations of motion for the action
\begin{align}
	S\ =\ & \frac{m}{4(m+1)} \int_{\bR\times M} \tr\left(\cf\La *\cf + 2\left(\frac{m}{m+1}\right)^2 \k d\t \La *_M Q_M \La\cf\La\cf\right) \nn \\ \vspace{10pt}
	  =\ & Vol(M)\times \int_\bR \left[ -\frac{1}{2}(\dot\chi^2+\dot\psi^2) - \left(\frac{m+1}{m}\right)^2  \right. \nn \\ 
		& \hspace{3cm} \left. \left( \psi^2 (1-\chi)^2 + m(1-m)(1-\k) \left(\frac{1}{2m}\psi^2-1\right)^2 \right. \right. \nn \\
			& \hspace{3.5cm} \left. \left. + m(1+\k(m-1)) \left(\chi-\frac{1}{2m}\psi^2\right)^2 \right) \right] d\t  \label{eq:YMaction}
\end{align}
with potential
\begin{align}
	V(\chi,\psi) = \frac{1}{2} &\left(\frac{m+1}{m}\right)^2 \left( ((1+\k(m-1)) m \chi^2 + \left( \k(1-m)-3 \right)\chi\psi^2 \right. \nn \\
		+ & \chi^2\psi^2 + \left.\left( 2-m+\k(m-1) \right)\psi^2 + \frac{1}{4} \psi^4 + m(m-1)(1-\k) \right), \label{eq:SEPotential}
\end{align}
where $*_M$ denotes the Hodge star operator on the Sasakian mainfold $M$ with respect to the metric $g_M=e^1 e^1 + \frac{2m}{m+1} \d_{ab} e^a e^b$, $*$ denotes the Hodge star operator on the cylinder, and $Vol(M)=\sqrt{|g_M|}e^{1,2,\cdots,2m+1}$ is the volume form on $M$. This can be verified by a direct computation, presented in Appendix \ref{app:Action}. Equations (\ref{eq:YMSE1}) constitute a gradient system of the form
\begin{align}
	\left(
	\begin{array}{c}
		\ddot \chi\\
		\ddot \psi
	\end{array}
	\right)
	=
	\left(
	\begin{array}{c}
		\p_\chi \\
		\p_\psi 
	\end{array}
	\right) V.
\end{align}
With our sign convention, this model describes a particle moving in the potential $-V$. The potential $V$ is symmetric with respect to sign changes of $\psi$ and has the following critical points (i.e. $\ddot \chi=\ddot\psi=0$) for arbitrary $m,\k$:
\begin{align}
	(\chi_1,\psi_1) = &(0,0), \nn\\
	(\chi_2,\psi_2) = &(1,\pm \sqrt{2 m}), \nn \\
	(\chi_3,\psi_3) = &\left(\frac{1}{4}\left(7+3(m-1)\k+\sqrt{P}\right), \right. \nn \\ 
		& \hspace{0.8cm} \pm \left. \frac{1}{2} \sqrt{\left((1-m)\k-1\right)\left((1-m)\k-1+4m+\sqrt{P}\right)} \right), \nn \\
	(\chi_4,\psi_4) = &\left(\frac{1}{4}\left(7+3(m-1)\k+\sqrt{P}\right), \right. \nn \\
		& \hspace{0.8cm} \left. \pm \frac{1}{2} \sqrt{\left((1-m)\k-1\right)\left((1-m)\k-1+4m-\sqrt{P}\right)} \right), \label{eq:SECritPoints}
\end{align}
where the abbreviation 
\begin{align}
	P=(m-1)^2\k^2+\k(8m^2-6m-2)+24m+1
\end{align}
is used. Finite-action Yang-Mills solutions $\chi(\t),\psi(\t)$ must interpolate between zero potential critical points. With $\k$ arbitrary, the potential vanishes for the second critical point $(\chi_2,\psi_2)=(1,\pm\sqrt{2m})$. We find $V(\chi_1,\psi_1)=\frac{(\k-1)(m-1)(m+1)^2}{2m}$ for the first critical point, which vanishes only for $\k=1$, as well as lengthy nonzero expressions for $V(\chi_3,\psi_3)$ and $V(\chi_4,\psi_4)$. The critical points are listed in Table \ref{tab:CritPts}, together with the $\k$-values for which their potential becomes zero.  
\begin{table}
\begin{centering}
\renewcommand{\arraystretch}{2.4}
\begin{tabular}{|m{1.5cm} m{0.1cm} m{2cm} c m{0.1cm} c|}
\hline
		 								 & 										 & & $\k$ 																& & Eigenvalues of Jacobian 					 \\ \hline \hline
	$(\chi_1,\psi_1) =$& $(0,0)$ 						 & & $1$					 												& & $(m+1)^2, \dfrac{(m+1)^2}{2}$			 \\ \hline
	$(\chi_2,\psi_2) =$& $(1,\pm\sqrt{2m})$	 & & any	 												 				& & see Appendix \ref{app:Eigenvalues} \\ \hline
	$(\chi_3,\psi_3) =$& $(1,-\sqrt{2m})$ 	 & & $\frac{m-2 - \sqrt{m(8+m)}}{2(m-1)}$ & & $0$, positive					 						 \\ \hline
	$(\chi_4,\psi_4) =$& $(-1,\pm\sqrt{2m})$ & & $\dfrac{3}{1-m}$ 										& & $\frac{(m+1)^2 \left(m-\sqrt{m (m+8)}\right)}{m^2}, \frac{(m+1)^2 \left(m+\sqrt{m (m+8)}\right)}{m^2}$ \\
		 								 & $(1,\sqrt{2m})$ 		 & & $\frac{m-2 + \sqrt{m(8+m)}}{2(m-1)}$ & & $0$, positive 										 \\ \hline
\end{tabular}
\caption{Critical points and corresponding $\k$ values with vanishing potential}
\label{tab:CritPts}
\end{centering}
\vspace{1cm}
\end{table}
For the values of $\k$ listed in Table \ref{tab:CritPtsSpecialKappa}, more than two critical points are located on the same axis, and hence the system may admit analytic solutions.
\begin{table}
\begin{centering}
\renewcommand{\arraystretch}{2.4}
\begin{tabular}{|c m{0.3cm} c m{0.3cm} c|}
\hline
	$\k$ &						 & Critical points &																					& $V($critical points$)$  \\ \hline \hline
	$\dfrac{1}{1-m}$ & & $(0,0), (1,\pm\sqrt{2m}), (1\pm m,0)$															& & $\dfrac{1}{2}(m+1)^2,0,\dfrac{1}{2}(m+1)^2$								 				\\ \hline
	\multirow{2}{*}{$\dfrac{3}{1-m}$} & & $(0,0), (1,\pm\sqrt{2m}), (-1,\pm\sqrt{2m}),$ & & \multirow{2}{*}{$\dfrac{(m+1)^2 (m+2)}{2 m},0,0,-\dfrac{(m+1)^2}{2 m^2}$} \\
	&  & $(0,\pm\sqrt{2(m+1)})$ & & \\
	\hline
\end{tabular} 
\caption{Values of $\k$ for which more than two critical points lie on the same axis}
\label{tab:CritPtsSpecialKappa}
\end{centering}
\vspace{1cm}
\end{table}
In addition, we note that at $\k=\frac{m-2}{m-1}$, five of the seven critical points coincide at $(0,0)$, at $\k=\frac{m-2 - \sqrt{m(8+m)}}{2(m-1)}$ the point $(\chi_3,\psi_3)$ coincides with $(\chi_2,\psi_2)$ and $(\chi_4,\psi_4)$ becomes imaginary, and at $\k=\frac{2-m-\sqrt{m(8+m)}}{2(m-1)}$, $(\chi_4,\psi_4)$ coincides with  $(\chi_2,\psi_2)$ and $(\chi_3,\psi_3)$ becomes imaginary.

\vspace{1cm}
\subsection{Analytic Yang-Mills solutions}
\label{ch:Solutions}
\begin{figure}[h]%
\begin{center}
	\includegraphics[width=7cm]{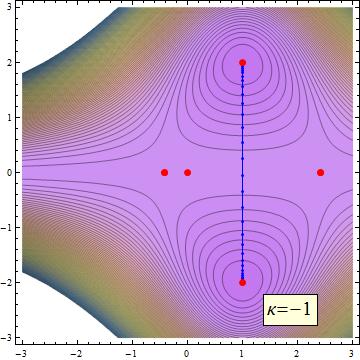} 
\end{center}
\caption{Analytic solution of the Yang-Mills equation for $\k=-1$ and $m=2$}%
\label{fig:Anal_Sol_m2}%
\end{figure}

Equations (\ref{eq:YMSE1}) constitute a system of nonlinear coupled differential equations, hence we cannot expect to be able to find analytic solutions. The case $\k=\frac{1}{1-m}$, however, admits an analytic solution to the Yang-Mills equation, interpolating between the critical points $(1,\sqrt{2m})$ and $(1,-\sqrt{2m})$ for arbitrary $m$. All other critical points are located on the $\chi$-axis and have potential $V=\frac{1}{2}(m+1)^2$. The zero-potential critical points are therefore minima of $V$, and we expect to find interpolating finite-action Yang-Mills solutions. With $\chi=1$, equations (\ref{eq:YMSE1}) take the form 
		\begin{subequations}
		\begin{align}
			\ddot\chi &= 0, \\
			\ddot\psi &= \frac{(m+1)^2}{m}\psi\left(\frac{1}{2m}\psi^2-1\right).
		\end{align} \label{eq:Psi_Chi1}
		\end{subequations}
		Equation (\ref{eq:Psi_Chi1}) is solved by
		\begin{align}
			\psi = \pm \sqrt{2m}\tanh \left(\pm\frac{m+1}{\sqrt{2m}}\t\right). \label{eq:KinkSol}
		\end{align}
		This is a kink solution with finite energy and finite action. A plot of this solution in the $(\chi,\psi)$-plane can be found in Figure \ref{fig:Anal_Sol_m2}.
		
		For $\k=\frac{3}{1-m}$, there are three critical points on the $\chi=0$ axis. However, none of them has zero potential, and we do not find any analytic solutions.

\vspace{1cm}
\subsection{Periodic solutions}
\label{ch:PeriodicSolutions}
A different kind of solutions is obtained by changing from $\bR\times M$ to $S^1\times M$, i.\ e.\ when the additional direction is not a real line but a unit circle with circumference $L$. In this case, periodic boundary conditions have to be imposed:
\begin{align}
	\psi(\t)=\psi(\t+L).
\end{align}
We restrict the consideration to the analytically solvable case (\ref{eq:Psi_Chi1}), which has the periodic solution
\begin{align}
	\psi(\t) = \pm \frac{2k\sqrt{m}}{\sqrt{1+k^2}} \text{ sn}\left[\frac{m+1}{\sqrt{m(1+k^2)}}\t;k\right]. \label{eq:PeriodicBoundary}
\end{align}
This solution is known as a sphaleron \cite{Manton2004}. Sphalerons are unstable solutions of the classical equations of motion. sn$[u,k]$ with $0\leq k\leq 1$ is a Jacobi elliptic function, details of which can be found for example in Appendix B of \cite{Bauer2010} or in \cite{Schwalm}. The Jacobi elliptic function has a period of $4K(k)$, where $K(k)$ denotes the complete elliptic integral of the first kind. The boundary condition (\ref{eq:PeriodicBoundary}) therefore turns into
\begin{align}
	4 K(k) n &= \frac{m+1}{\sqrt{m(1+k^2)}}L, && n\in\bN,
\end{align}
fixing $k=k(L,n)$ and $\psi(\t;k(L,n))=:\psi^{(n)}(\t)$. Solutions (\ref{eq:PeriodicBoundary}) exist if $L\geq 2^{\frac{3}{2}} \pi n$ (cf. \cite{Rahn2009,Bauer2010}). The topological charge of the sphaleron $\psi{(n)}$ is zero due to the periodic boundary conditions. This solution is interpreted as a configuration of $n$ kinks and $n$ antikinks, alternating and equally spaced around the circle. The $\tanh$-solution from Chapter \ref{ch:Solutions} arises from the Jacobi elliptic function in the limit $k\ra 1$. In the limit $k\ra 0$, the elliptic function approaches $\sin\left(\frac{m+1}{\sqrt{m(1+k^2)}}\t\right)$. In analogy to results in \cite{Manton1988}, our solution (\ref{eq:PeriodicBoundary}) with positive sign has the following total energy, with $E(k)$ denoting the complete elliptic integral of the second kind:
\begin{align}
 E[\psi] =\ &\int_0^L d\t \left( \frac{1}{2}(\p_\t\psi)^2 + V(1,\psi)\right) \nn \\
  =\ &\frac{\sqrt 2 \cdot 4 n m^2 (m+1)}{3(1+k²)^{\frac{3}{2}}} \nn \\ 
    & \hspace{0.5cm} \left( \frac{1}{4m^2} \left( 3k^4+(6+32m^2+24m)k^2+16m^2-24m+3 \right) K(k) \right. \nn \\
     & \hspace{1cm} + \left. 2\left(\frac{3}{m}-2\right)(1+k^2) E(k) \right).
\end{align}

\vspace{1cm}
\subsection{Dyons}
\label{ch:DyonSolutions}
Replacing the coordinate $\t$ in $\bR$ direction by $i\t$ changes the signature of the metric from Riemannian to Lorentzian:
\begin{align}
	g &= -e^0 e^0 + e^1 e^1 + e^{2h} \d_{ab} e^{ab}.
\end{align}
The Yang-Mills equations (\ref{eq:YMSE1}) remain unchanged, except for the fact that the second-order derivatives now come with a minus sign:
\begin{align}
	(\ddot \chi,\ddot \psi) \ra (-\ddot \chi,-\ddot \psi).
\end{align}
This corresponds to a sign flip of the potential, so that we have to study $V$ instead of $-V$. Dyons are finite-energy solutions to the second-order equations obtained by this sign flip. Just as Yang-Mills solutions, they can interpolate between two critical points (kink), or start and end at the same point (bounce). Solutions that oscillate around a minimum can exist as well, but they do not lead to finite energy and hence will not be considered in the following.

\vspace{1cm}
\subsection{Discussion and summary of solutions}
Recall that in our sign convention, instanton solutions interpolate between minima and dyon solutions between maxima of $V$. In both cases, solutions that start or end at a saddle point are possible as well.  With this in mind, we can expect the following solutions:
\begin{itemize}
	\item $\k$ arbitrary: there exist at least two zero-potential critical points at $(0,\pm\sqrt{2m})$ for all $\k$. According to Appendix \ref{app:Eigenvalues}, they can be minima or saddle points of $V$, depending on the value of $\k$. This means that we can always find interpolating solutions, either of dyon or of Yang-Mills type. These solutions have to be constructed numerically unless $\k=\frac{1}{1-m}$.
	\item $\k=1$: this is the instanton case. Yang-Mills solutions exist between $(0,0)$ and $(1,\pm\sqrt{2m})$ (cf. \cite{Harland2011}). We do not expect to find any finite-action dyon solutions, as the zero-potential critical points of $V$ are minima. 
	\item $\k=\frac{1}{1-m}$ ($\k=-1$ for $m=2$): in this case, we find three nonzero critical points along the $\chi$ axis. An analytic Yang-Mills solution interpolates between the two remaining zero-potential critical points, which are minima for all $m$. This solution for arbitrary $m$ is presented in Chapter \ref{ch:Solutions}.
	\item $\k=\frac{3}{1-m}$ ($\k=-3$ for $m=2$): we find four zero-potential critical points. Two of them are located at the lines with $\chi=1$ and $\chi=-1$, respectively. We do not find any analytic solutions along the $\chi=\pm 1$ and $\chi=0$ axes. There should, however, be a number of numerical solutions interpolating between various pairs of critical points. 
\end{itemize}

\begin{figure}[h]%
	\includegraphics[width=7cm]{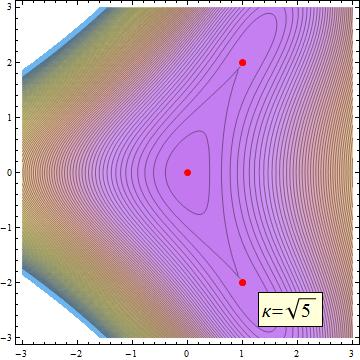} \hspace{2cm}
	\includegraphics[width=7cm]{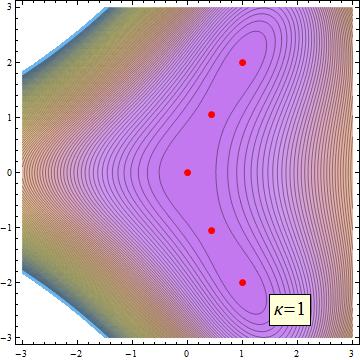} \vspace{0.5cm} \\
	\includegraphics[width=7cm]{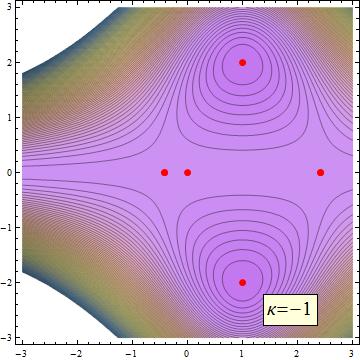} \hspace{2cm} 
	\includegraphics[width=7cm]{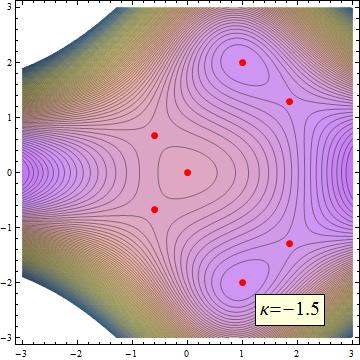} \vspace{0.5cm} \\
	\includegraphics[width=7cm]{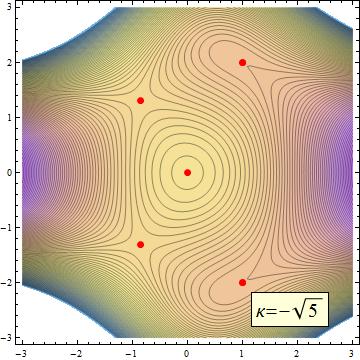} \hspace{2cm} 
	\includegraphics[width=7cm]{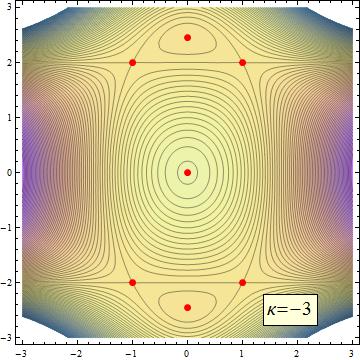} 
\caption{Potential plots for various values of $\k$ and $m=2$}%
\label{fig:Pot_m2}%
\end{figure}

\begin{figure}[h]%
	\includegraphics[width=7cm]{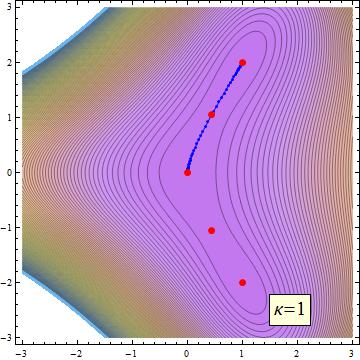} \hspace{2cm}
	\includegraphics[width=7cm]{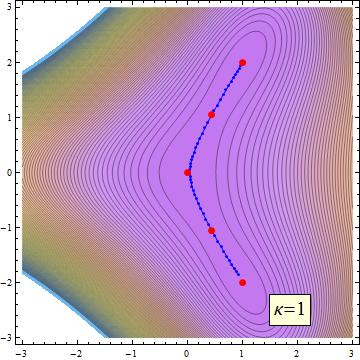} \vspace{0.5cm} \\
	\includegraphics[width=7cm]{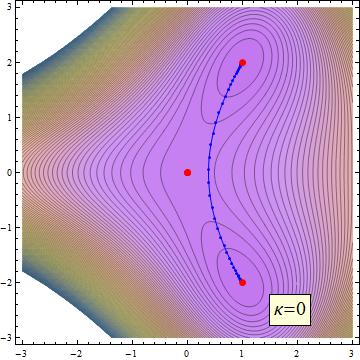} \hspace{2cm}
	\includegraphics[width=7cm]{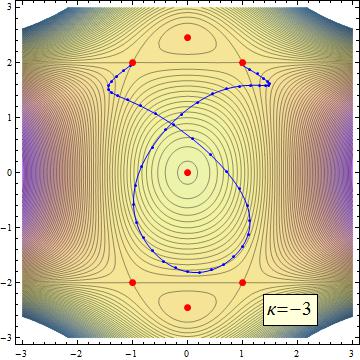} \vspace{0.5cm} \\
	\includegraphics[width=7cm]{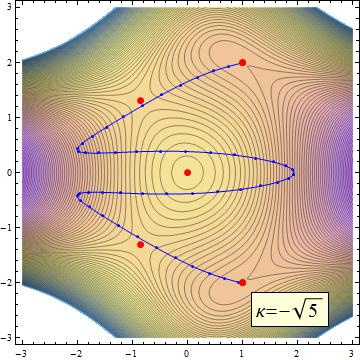} \hspace{2cm}
	\includegraphics[width=7cm]{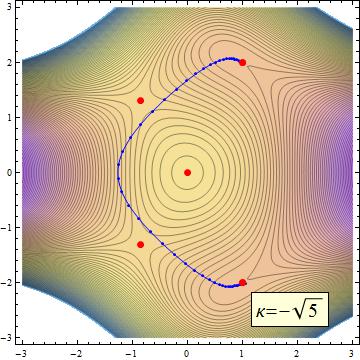} 
\caption{Some solutions of the Yang-Mills equation for various values of $\k$ and $m=2$}%
\label{fig:Num_Sol_YM_m2}%
\end{figure}

\begin{figure}[h]%
	\includegraphics[width=7cm]{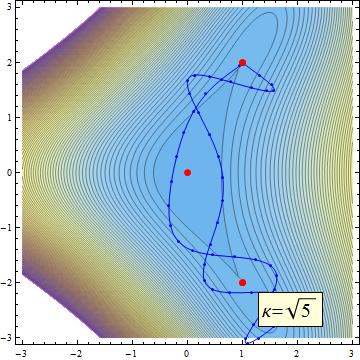} \hspace{1cm}
	\includegraphics[width=7cm]{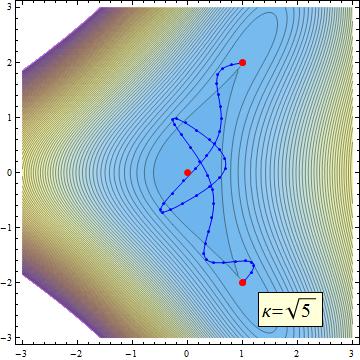} \vspace{0.5cm} \\
\begin{center}
	\includegraphics[width=7cm]{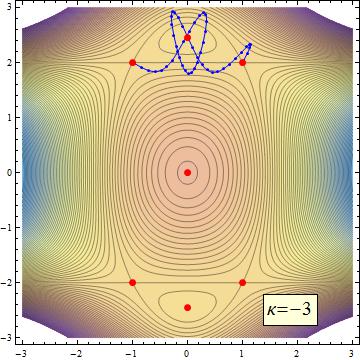}
\end{center}
\caption{Some numerical dyon solutions for various values of $\k$ and $m=2$}%
\label{fig:Num_Sol_Dyon_m2}%
\end{figure}

We do not expect any analytic dyon solutions, as the zero-potential critical points are minima in the analytically solvable cases. For a better understanding, we present the case $m=2$ as an example. The potential for various interesting values of $\k$ is shown in Figure \ref{fig:Pot_m2}, and further dyon and Yang-Mills solutions for this example are presented in Figures \ref{fig:Num_Sol_YM_m2} and \ref{fig:Num_Sol_Dyon_m2}. The list of zero-potential critical points can be found in Table \ref{tab:CritPtsm2}.

\begin{table}
\begin{centering}
\renewcommand{\arraystretch}{2.4}
\begin{tabular}{|m{1.5cm} m{1.8cm} c m{0.3cm} c|}
\hline
		 							 &  				 & $\k$ 				& & Eigenvalues of Jacobian \\ \hline \hline
	$(\chi_1,\psi_1) =$& $(0,0)$		 & $1$ 					& & $9,\dfrac{9}{4}$ 								 \\ \hline 
	$(\chi_2,\psi_2) =$& $(1,\pm2)$	 & any	& & $\dfrac{9}{4}\left(5+\k+\sqrt{5}(1+\k)\right),\dfrac{9}{4} \left(5+\k-\sqrt{5}(1+\k)\right)$ \\ \hline
	$(\chi_3,\psi_3) =$& $(1,-2)$		 & $-\sqrt 5 	$ & & $\dfrac{9}{2} \left(5-\sqrt{5}\right),0$\\ \hline
	$(\chi_4,\psi_4) =$& $(-1,\pm2)$ & $-3$					& & $\dfrac{9}{2} \left(1+\sqrt{5}\right),\dfrac{9}{2} \left(1-\sqrt{5}\right)$ \\
		 							 & $(1,2)$		 & $\sqrt 5$ 		& & $\dfrac{9}{2} \left(5+\sqrt{5}\right),0$ \\ \hline
\end{tabular}
\caption{Critical points and corresponding $\k$ values with vanishing potential for $m=2$}
\label{tab:CritPtsm2}
\end{centering}
\vspace{1cm}
\end{table}
\bigskip

\newpage
\section{Conclusion and outlook}
Using a special ansatz for the gauge connection, we have derived a system of explicit second-order Yang-Mills equations on the cylinder over a class of Sasakian manifolds. We have constructed the corresponding action and potential, discussed the behaviour of the critical zero-potential points and found analytic as well as numerical solutions of Yang-Mills, dyon and sphaleron type. 
\bigskip

A similar discussion for cylinders over certain $SU(3)$-structure manifolds can be found in \cite{Bauer2010}. A comparison with our results illustrates that Sasakian and $SU(3)$-structures are fundamentally different. The perhaps most striking fact is that the 3-symmetry of the $SU(3)$-structure manifold is recovered in the shape of the potential, whereas the potential in the Sasakian case is symmetric only under sign changes of the variable $\psi$. Furthermore, the Sasakian potential does not admit as many solutions with straight trajectories in the $(\chi,\psi)$-plane as the $SU(3)$-structure potential does. In the latter case, the distribution of $\k$-dependent and $\k$-independent zero-potential critical points allows to systematically associate certain types of solutions (kinks, bounces) to intervals of the deformation parameter $\k$. In particular, there are always three critical points on the real axis. The Sasakian potential admits fewer $\k$-independent zero-potential critical points, and they are not as regularly distributed as in the $SU(3)$-structure case. The range and type of our solutions is therefore significantly different. 
\bigskip

In spite of these differences, we have found that Sasakian manifolds do admit various interesting solutions. This, and in particular the fact that we have found an analytic kink-type solution of the Yang-Mills equation, makes them potentially interesting for non-supersymmetric string compactifications. It may be worth studying the instanton solution (\ref{eq:KinkSol}) in the context of the AdS/CFT duality mentioned in the introduction. 
\bigskip

To complete the discussion, it would be interesting to consider Yang-Mills and dyon solutions on cylinders over $G_2$-structure manifolds, i.\ e.\ 8-dimensional manifolds with $Spin(7)$-structure. We are planning to present results for this case in the near future. In addition, the analysis for the 3-Sasakian case is still missing. Another open question is how the Yang-Mills and dyon solutions change when considering cones and sine-cones instead of cylinders. On conical manifolds, the second-order equations acquire a first-order friction term, hence the analysis might have to be done numerically.

\vspace{1cm}
\appendix
\section{Sum of structure constants} \label{sec:beta}
\bp
Let $M$ be a Sasakian manifold of dimension $2m+1$ with structure group $SU(m)$ and metric 
\begin{align}
	g_M &= e^1 e^1 + \frac{2m}{m+1} \d_{ab} e^{ab}. \label{eq:SEMetric2}
\end{align}
Let $\bR\times M$ be the cylinder with structure group $SU(m+1)$. Then the Lie algebras corresponding to the structure groups admit a splitting $\gs\gu(m+1)=\gs\gu(m)\oplus \gm$, as described in Chapter \ref{ch:YMonSE}. We use indices $a=(1,2,\ldots, \dim \gm)$ to label the generators of $\gm$ and indices $i,j$ for the remaining generators of $\gs\gu(m)$. In this setup, the $SU(m+1)$-structure constants satisfy equation (\ref{eq:beta}):
\begin{align}
	f_{ac}^i f_{ib}^c &= \frac{2 (m^2-1)}{m} \d_{ab}. \label{eq:SasakiStrConst}
\end{align}
\ep

\pf To see this, note first that the components of the metric (\ref{eq:SEMetric2}) take the form 
\begin{align}
	\left(g_M\right)_{11} &= 1, && \left(g_M\right)_{ab} = \frac{2m}{m+1} \d_{ab}.
\end{align}
The Killing form of $\gs\gu(m+1)$ induces the following metric on $\gm$:
\begin{align}
	\left(g_{K}\right)_{\mu\nu} &= f_{\mu \wt c}^{\wt d} f_{\wt d \nu}^{\wt c}.
\end{align}
The structure constants are normalized such that they satisfy
\begin{align}
	f_{ab}^1 &= 2 P_{ab1}, && f_{1a}^b = \frac{m+1}{m} P_{1ab}.
\end{align}
Hence, the Killing metric takes the following values:
\begin{align}
	\left(g_{K}\right)_{11} &= f_{1\wt c}^{\wt d} f_{\wt d1}^{\wt c} = f_{1c}^d f_{d1}^c = \frac{2(m+1)^2}{m} =: X, \\
	\left(g_{K}\right)_{ab} &= 2 (f_{a1}^d f_{db}^1+f_{ai}^d f_{db}^i) = 2\left(\frac{2(m+1)}{m} \d_{ab} + f_{ai}^d f_{db}^i\right).
\end{align}
This metric matches the metric (\ref{eq:SEMetric2}) up to rescaling of structure constants by the factor $\sqrt{X}$. We therefore find
\begin{align}
	(g_M)_{11} &= \frac{1}{X} \left(g_{K}\right)_{11} = 1, \\
	(g_M)_{ab} &= \frac{1}{X} \left(g_{K}\right)_{ab} \nn \\
		&= \frac{2}{X} (f_{a1}^d f_{db}^1+f_{ai}^d f_{db}^i) \nn \\
		&= \frac{2}{X} \left( \frac{2(m+1)}{m} \d_{ab} + f_{ai}^d f_{db}^i \right) \nn \\
		&= \frac{2m}{m+1} \d_{ab}.
\end{align}
We conclude that both summands in $(g_M)_{ab}$ must be proportional to $\d_{ab}$, hence $f_{ai}^d f_{db}^i\stackrel{!}{=}\b\d_{ab}$ with some real parameter $\b\in \bR$. This leads to
\begin{align}
	\frac{2}{X}\left(\frac{2(m+1)}{m}+\b\right) \d_{ab} &= \frac{2m}{m+1} \d_{ab} && \Ra && \b = \frac{2(m^2-1)}{m}
\end{align}
and proves equation (\ref{eq:SasakiStrConst}).
\qed

\vspace{1cm}
\section{Action}
\label{app:Action}
\bp
The Yang-Mills equations (\ref{eq:YMSE1}) on the cylinder over a Sasakian manifold are equations of motion for the action
\begin{align}
	S\ =\ & \frac{m}{4(m+1)} \int_{\bR\times M} \tr\left(\cf\La *\cf + 2\left(\frac{m}{m+1}\right)^2 \k d\t \La *_M Q_M \La\cf\La\cf\right) \nn \\ \vspace{10pt}
	  =\ & Vol(M)\times \int_\bR \left[ -\frac{1}{2}(\dot\chi^2+\dot\psi^2) - \left(\frac{m+1}{m}\right)^2  \right. \nn \\ 
		& \hspace{3cm} \left. \left( \psi^2 (1-\chi)^2 + m(1-m)(1-\k) \left(\frac{1}{2m}\psi^2-1\right)^2 \right. \right. \nn \\
			& \hspace{3.5cm} \left. \left. + m(1+\k(m-1)) \left(\chi-\frac{1}{2m}\psi^2\right)^2 \right) \right] d\t \label{eq:YMactionApp}
\end{align}
with potential
\begin{align}
	V(\chi,\psi) = \frac{1}{2} &\left(\frac{m+1}{m}\right)^2 \left( ((1+\k(m-1)) m \chi^2 + \left( \k(1-m)-3 \right)\chi\psi^2 \right. \nn \\
		+ & \chi^2\psi^2 + \left.\left( 2-m+\k(m-1) \right)\psi^2 + \frac{1}{4} \psi^4 + m(m-1)(1-\k) \right), 
\end{align}
where $*_M$ denotes the Hodge star operator on the Sasakian manifold $M$ with respect to the metric $g_M$, and $*$ denotes the Hodge star operator on the cylinder. 
\ep

\pf
To see this, we compute the summands $\tr(\cf\La *\cf)$ and $\tr(d\t \La *_M Q_M \La\cf\La\cf)$ separately. For the first summand, we find
\begin{align}
	\tr(\cf\La *\cf) =& \frac{1}{2}\tr(2\cf_{0\mu}\cf^{0\mu} + \cf_{\mu\nu}\cf^{\mu\nu}) Vol(\bR\times M) \nn \\
		= & \frac{1}{2}\tr\left(2\cf_{01}\cf_{01} + \frac{m+1}{m} \cf_{0a}\cf_{0a} \right. \nn \\
			& \hspace{2cm} \left. + \frac{m+1}{m} \cf_{1b}\cf_{1b} + \left(\frac{m+1}{2m}\right)^2 \cf_{ab}\cf_{ab}\right) Vol(\bR\times M) \nn \\
		= &\frac{1}{2}\left(-4 \frac{m+1}{m} \left(\dot\chi^2+\dot\psi^2\right) -4 \left( \frac{m+1}{m}\right)^3 \psi^2(1-\chi)^2 \right. \nn \\
		& \hspace{2cm} \left. + \left( \frac{m+1}{2m}\right)^2 \left( \left( \frac{1}{2m}\psi^2-1 \right)^2 f_{ab}^i f_{ab}^j f_{im}^n f_{jn}^m \right. \right. \nn \\
		& \hspace{2cm} \left. \left. - 16(m+1) \left( \chi - \frac{1}{2m}\psi^2 \right)^2 \right) \right) Vol(\bR\times M), \label{eq:action1}
\end{align}
using $Vol(\bR\times M)=\sqrt{|g_\cz|}d\t\La e^{1\cdots (2m+1)}$, the components (\ref{eq:FSasaki}) of the curvature and the following explicit expressions for the trace of $I_i,I_\mu$ in the representation (\ref{eq:Generators}):
\begin{align}
	\tr(I_1 I_1) &= I_{10}^a I_{1a}^0 + I_{1a}^b I_{1b}^a + I_{11}^0 I_{10}^1 + I_{10}^1 I_{11}^0 = -2 \frac{m+1}{m}, \\
	\tr(I_i I_j) &= I_{ia}^b I_{jb}^a, \\
	\tr(I_1 I_j) &= 0, \\
	\tr(I_a I_a) &= 2(I_{a0}^b I_{ab}^0 + I_{a1}^b I_{ab}^1) = -8m \text{ (sum over $a$)}.
\end{align}
The combination $f_{ab}^i f_{ab}^j f_{im}^n f_{jn}^m$ of structure constants in equation (\ref{eq:action1}) can be simplified by use of the following relation. The commutator of two generators in the representation (\ref{eq:Generators}) takes the form 
\begin{align}
	[I_a,I_b]_c^d = f_{ab}^i I_{ic}^d + f_{ab}^1 I_{1c}^d.
\end{align}
Inserting the explicit expressions for $I_1$ and $I_a$ leads to the identity
\begin{align}
	f_{ab}^i f_{ic}^d = \o_{bc}\o_{ad} - \o_{ac}\o_{bd} - \d_a^c\d_b^d + \d_b^c\d_a^d + \frac{2}{m} P_{ab1} \o_{cd}. \label{eq:StrConstfff}
\end{align}
We use this expression to rewrite the sum of structure constants in equation (\ref{eq:action1}) and find 
\begin{align}
 \sum_{a,b,c,d,i,j} f_{ab}^i f_{ab}^j f_{im}^n f_{jn}^m &= \sum_{a,b,c,d,i,j} \left(\o_{bc}\o_{ad} - \o_{ac}\o_{bd} - \d_a^c\d_b^d + \d_b^c\d_a^d + \frac{2}{m} \o_{ab} \o_{cd}\right) \nn \\
		&\hspace{2.5cm} \left(\o_{bd}\o_{ac} - \o_{ad}\o_{bc} - \d_a^d\d_b^c + \d_b^d\d_a^c - \frac{2}{m} \o_{ab} \o_{cd}\right) \nn \\
		&= \sum_{a,b,c,d,i,j} \left(2 \o_{bc}\o_{ad}\o_{bd}\o_{ac} - 2\o_{ac}\o_{bd}\o_{ac}\o_{bd} + \frac{4}{m} \o_{ab}\o_{cd}\o_{bd}\o_{ac} \right. \nn \\
		&\hspace{2.5cm} \left.- \frac{4}{m} \o_{ab}\o_{cd}\o_{bc}\o_{ad} -\frac{4}{m^2}\o_{ab}\o_{cd}\o_{ab}\o_{cd} \right. \nn \\
		&\hspace{2.5cm} + \left. \left(\frac{8}{m}-4\right)\o_{ab}\o_{ab} + 2(\d_a^d\d_a^d-\d_a^d\d_b^c) \right) \nn \\
		&= 4m-8m^2+8+8-16+\left(\frac{8}{m}-4\right)2m+4m-8m^2 \nn\\
		&= 16 (1-m^2), \label{eq:fff1}
\end{align}
using the fact that $\o_{ab}\o_{ab}=2m$ and that only the components of $\o_{ab}$ with $b=a+1$ or $b=a-1$ are nonzero. To avoid confusion, the summation indices have been explicitly displayed at this point. Note that all indices are being summed over. Inserting this back into equation (\ref{eq:action1}) yields
\begin{align}
	\tr(\cf\La *\cf) = &4\frac{m+1}{m} \left( -\frac{1}{2} \left(\dot\chi^2+\dot\psi^2\right) - \left( \frac{m+1}{m}\right)^2 \psi^2(1-\chi)^2 \right. \nn \\
			&\hspace{2cm} + \left. (1-m^2) \frac{m+1}{m} \left( \frac{1}{2m}\psi^2-1 \right)^2 \right. \nn \\
			&\hspace{2cm} - \left. \frac{(m+1)^2}{m} \left( \chi - \frac{1}{2m}\psi^2 \right)^2 \right) Vol(\bR\times M) \nn \\
		= &4\frac{m+1}{m} \left( -\frac{1}{2} \left(\dot\chi^2+\dot\psi^2\right) - \left( \frac{m+1}{m}\right)^2 \left( \psi^2(1-\chi)^2 \right. \right. \nn \\
		&\hspace{2cm} - \left. \left. (1-m)m \left( \frac{1}{2m}\psi^2-1 \right)^2 \right.\right. \nn \\
		&\hspace{2cm} + \left.\left. m \left( \chi - \frac{1}{2m}\psi^2 \right)^2 \right) \right) Vol(\bR\times M). \label{eq:action1a}
\end{align}

For the second summand in the action, note that $d\t\La *_M Q_M\La \cf\La \cf$ is a form of top degree in on the cylinder $\cz(M)$. A convenient way to compute the components of this form is to apply the Hodge star operator. We find
\begin{align}
	*_{\cz(M)}(d\t\La *_M Q_M\La \cf\La \cf) &= *_M(*_M Q_M\La \cf\La \cf) \nn \\
		&= \frac{1}{4 n! (n-4)!} Q_{\mu\nu\r\s} \cf^{\a\b} \cf^{\g\d} \e^{\mu\nu\r\s\xi_1 \cdots \xi_{n-4}} \e_{\xi_1 \cdots \xi_{n-4} \a\b\g\d} \nn \\
		&= \frac{1}{4} Q_{\mu\nu\r\s} \cf^{\a\b} \cf^{\g\d} \d_{\a\b\g\d}^{\mu\nu\r\s} \nn \\
		&= \frac{1}{4} Q_{\mu\nu\r\s} \cf^{[\mu\nu} \cf^{\r\s]} \nn \\
		&= 3 \o_{\mu\nu} \o_{\r\s} \cf^{[\mu\nu} \cf^{\r\s]}, 
\end{align}
using $n:=2m+1=\dim M$ as well as $Q=\frac{1}{2}\o\La\o\LRa Q_{\mu\nu\r\s}=4!\o_{\mu\nu}\o_{\r\s}$. This result implies 
\begin{align}
	d\t\La *_M Q_M\La \cf\La \cf &= 3\o_{ab}\o_{cd} \cf^{[ab} \cf^{c]d} Vol(\bR\times M).
\end{align}
We find
\begin{align}
	\tr(d\t\La *_M& Q_M\La \cf\La \cf) \nn \\
		&= 3\o_{ab}\o_{cd} \tr(\cf^{[ab} \cf^{c]d}) Vol(\bR\times M) \nn \\
		&= \left(\frac{m+1}{2m}\right)^4 \left( \left( \frac{1}{2m}\psi^2-1 \right)^2 3 \o_{ab}\o_{cd} f_{[ab}^i f_{c]d}^j f_{im}^n f_{jn}^m \right. \nn \\
			&\hspace{1.5cm}- \left. 8 \frac{m+1}{m} \left( \chi - \frac{1}{2m}\psi^2 \right)^2 3 \o_{ab} \o_{cd} P_{[ab|1|} P_{c]d1} \right) Vol(\bR\times M) \nn \\
		&= \left(\frac{m+1}{2m}\right)^4 \left( 2 \left( \frac{1}{2m}\psi^2-1 \right)^2 \o_{ab}\o_{cd} f_{bc}^i f_{ad}^j f_{im}^n f_{jn}^m \right. \nn \\
			&\hspace{1.5cm}- \left. 32 (m^2-1) \left( \chi - \frac{1}{2m}\psi^2 \right)^2 \right) Vol(\bR\times M),
\end{align}
by use of $f_{ab}^1f_{ab}^i=0$ and $3 \o_{ab} \o_{cd} P_{[ab|1|} P_{c]d1} = 4m(m-1)$. The sum of structure constants simplifies to 
\begin{align}
	\o_{ab}\o_{cd} f_{[ab}^i f_{c]d}^j f_{im}^n f_{jn}^m = 16 (m^2-1).
\end{align}
This identity is proven by writing the structure constants in terms of equation (\ref{eq:StrConstfff}) and evaluating all sums explicitly. As the computation follows the same pattern as the derivation of equation (\ref{eq:fff1}), we do not present the details here. We find
\begin{align}
	\tr(d\t&\La *_M Q_M\La \cf\La \cf) \nn \\ 
		&= \left(\frac{m+1}{2m}\right)^4 32 (m^2-1) \left( \left( \frac{1}{2m}\psi^2-1 \right)^2 - \left( \chi - \frac{1}{2m}\psi^2 \right)^2 \right) Vol(\bR\times M). \label{eq:action2}
\end{align}
Now the identities (\ref{eq:action1a}) and (\ref{eq:action2}) can be inserted into the action. This leads to the result (\ref{eq:YMactionApp}), taking into account that the volume form on the cylinder satisfies ${Vol(\bR\times M)=d\t\La Vol(M)}$.
\qed

\vspace{1cm}
\section{Eigenvalues of the Hesse matrix}
\label{app:Eigenvalues}
Let us once again consider the potential (\ref{eq:SEPotential}):
\begin{align}
	V(\chi,\psi) = \frac{1}{2} &\left(\frac{m+1}{m}\right)^2 \left( ((1+\k(m-1)) m \chi^2 + \left( \k(1-m)-3 \right)\chi\psi^2 \right. \nn \\
		+ & \chi^2\psi^2 + \left.\left( 2-m+\k(m-1) \right)\psi^2 + \frac{1}{4} \psi^4 + m(m-1)(1-\k) \right).
\end{align}
The critical points $(\chi,\psi)$ of $V$ that satisfy $\p_\chi V=\p_\psi V=0$ are listed in equation (\ref{eq:SECritPoints}), and the eigenvalues of the matrix 
\begin{align}
	\left(
	\begin{array}{cc}
		\dfrac{\p^2 V}{\p \chi^2} & \dfrac{\p^2 V}{\p \chi \p\psi} \vspace{0.3cm} \\
		\dfrac{\p^2 V}{\p \psi \p\chi} & \dfrac{\p^2 V}{\p \psi^2}
	\end{array} 
	\right)
\end{align}
have been presented in Table \ref{tab:CritPts}. The eigenvalues at the critical point $(\chi_2,\psi_2) = (1,\pm\sqrt{2m})$ need a more detailed discussion. They are given by 
\begin{align}
	(\l_1,\l_2) = &\left(\frac{1}{2}\left(\frac{m+1}{m}\right)^2 \left( (5+\k(m+1))m + (1+\k(m-1))\sqrt{m(8+m)} \right),\right. \nn\\
								&\left.\frac{1}{2}\left(\frac{m+1}{m}\right)^2 \left( (5+\k(m-1))m - (1+\k(m-1))\sqrt{m(8+m)} \right)\right).
\end{align}
$\l_1$ is greater than zero for 
\begin{align}
	\k > \k_+ := -\frac{5m+\sqrt{m(8+m)}}{m(m+1)+(m-1)\sqrt{m(8+m)}} 
\end{align}
and smaller than zero otherwise. $\l_2$ is greater than zero for 
\begin{align}
	\k < \k_- := \frac{-5m+\sqrt{m(8+m)}}{m(m-1)-(m-1)\sqrt{m(8+m)}} 
\end{align}
and smaller otherwise. We have $\k_- > \k_+$ for any positive integer value of $m>1$. The extremum of the potential at $(1,\pm\sqrt{2m})$ is therefore
\begin{align}
\begin{array}{cccc}
	1) & \text{a saddle} 		& \text{ for } & \k > \k_-, \\
	2) & \text{indefinite} & \text{ for } & \k = \k_-, \\
	3) & \text{a minimum}  & \text{ for } & \k_- > \k > \k_+, \\
	4) & \text{indefinite} & \text{ for } & \k = \k_+, \\
	5) & \text{a saddle} 		& \text{ for } & \k_+ > \k. 
\end{array}
\end{align}
This observation is in agreement with the remaining cases listed in Table \ref{tab:CritPts}: since \linebreak $\k_+>\frac{3}{1-m}$, we find one positive and one negative eigenvalue for $(\chi_4,\psi_4)$. 

We can expect Yang-Mills solutions when the extrema at $(1,\pm\sqrt{2m})$ are minima, i.\ e.\  for $\k_->\k>\k_+$ (in particular for $\k=1$), or saddle points, and dyon solutions when they are saddle points. As $\l_1$ and $\l_2$ do not simultaneously become smaller than zero for any fixed value of $\k$, the critical points never become maxima.


\vspace{1cm}
\subsection*{Acknowledgements}
I am grateful to Derek Harland, Alexander Haupt and Olaf Lechtenfeld for discussions, advice and support. I also wish to thank Jian Qiu for helpful comments on the four-form $Q$. This work was partially supported by the Deutsche Forschungsgemeinschaft Research Training Group GRK 1463 and grant LE 838/13.



\end{document}

%% file: Pre.tex
\renewcommand{\a}{\alpha}  
\renewcommand{\b}{\beta}  
  
\renewcommand{\d}{\delta}  
\newcommand{\e}{\epsilon}  
  
\newcommand{\g}{\gamma}  
\newcommand{\h}{\eta}

\renewcommand{\k}{\kappa}  
\renewcommand{\l}{\lambda}  
\renewcommand{\o}{\omega}  
  
\renewcommand{\r}{\rho}  
\newcommand{\s}{\sigma}  
\renewcommand{\t}{\tau}

\newcommand{\G}{\Gamma}

\renewcommand{\O}{\Omega}

\newcommand{\bR}{\mathbb{R}}

\newcommand{\bN}{\mathbb{N}}

\newcommand{\gm}{\mathfrak{m}}

\newcommand{\gs}{\mathfrak{s}}  
  
\newcommand{\gu}{\mathfrak{u}}

\newcommand{\id}   {{\mathbbm{1}}}

\newcommand{\ca}{\mathcal{A}}

\newcommand{\cd}{\mathcal{D}}  
  
\newcommand{\cf}{\mathcal{F}}  
  
\newcommand{\ch}{\mathcal{H}}

\newcommand{\cn}{\mathcal{N}}

\newcommand{\cz}{\mathcal{Z}}

\newcommand{\p}{\partial}

\renewcommand{\square}{\kern1pt\vbox  
               {\hrule height 0.6pt\hbox{\vrule width 0.6pt\hskip 3pt  
    \vbox{\vskip 6pt}\hskip 3pt\vrule width 0.6pt}\hrule height0.6pt}  
    \kern1pt}  
  
\newcommand{\La}{\wedge}  
\newcommand{\ra}{\rightarrow}
\newcommand{\Ra}{\Rightarrow}
	\newcommand{\LRa}{\Leftrightarrow}

\newcommand{\nn}{\nonumber}

\DeclareMathOperator\tr{tr\;}

\newcommand{\wt}{\widetilde}

\newtheorem{Th}{Theorem}  
\newtheorem{Prop}{Proposition}  
  
\newtheorem{Cor}{Corrollary}  
\newtheorem{Lem}{Lemma}  
\newtheorem{Def}{Definition}  
\newcommand{\bt}{\begin{Th}\ \ }  
\newcommand{\et}{\end{Th}} 

\newcommand{\bp}{\begin{Prop}\ \ }  
\newcommand{\ep}{\end{Prop}}  
\newcommand{\bc}{\begin{Cor}\ \ }  
\newcommand{\ec}{\end{Cor}}  
\newcommand{\bl}{\begin{Lem}\ \ }  
\newcommand{\el}{\end{Lem}}  
\newcommand{\bd}{\begin{Def}\ \ }  
\newcommand{\ed}{\end{Def}}  
  
\newcommand{\pf}{\noindent{\it Proof:\ \ }}  

\newcommand{\n}{\nabla}

\newcommand{\be}{\begin{equation}}  
\newcommand{\ee}{\end{equation}}

\newcommand{\arr}{\begin{array}{rlll}}  
\newcommand{\ea}{\end{array}}  
\newcommand{\bea}{\begin{eqnarray}}  
\newcommand{\eea}{\end{eqnarray}}  
\newcommand{\bean}{\begin{eqnarray*}}  
\newcommand{\eean}{\end{eqnarray*}}  

